# PHOTOVOLTAIC EFFECT ON MOLECULE COUPLED FERROMAGNETIC FILMS OF A MAGNETIC TUNNEL JUNCTION

Pawan Tyagi[1,2], Christopher Riso[1]

[1]Mechanical Engineering, University of the District of Columbia, Washington, DC, 20008, USA
[2]Chemical and Materials Engineering, University of Kentucky, Lexington, Kentucky-40506, USA
Email:ptyagi@udc.edu

**ABSTRACT:**

Economical solar energy conversion to electricity can be boosted by the discovery of fundamentally new photovoltaic mechanism, and a suitable system to realize it with commonly available materials like iron (Fe) and nickel (Ni). This paper reports the observation of the photovoltaic effect on a molecular spintronics device, composed of a magnetic tunnel junction (MTJ) and organometallic molecular clusters (OMCs). A prefabricated MTJ with exposed side edges, after enabling the bridging of OMC channels between its two ferromagnetic films, exhibited following phenomenon (i) a dramatic increase in exchange coupling, (ii) 3-6 orders current suppression and (iii) photovoltaic effect. This paper focuses on the photovoltaic effect. Control experiments on isolated ferromagnetic films suggested that OMCs neither affected the magnetic properties nor produced any photovoltaic effect; the photovoltaic effect was only observed on the ferromagnetic films serving as magnetic electrodes in a MTJ. Present paper invites further investigation of the similar photovoltaic effect on other combinations of MTJs and promising magnetic molecules, like single molecular magnets, organometallic clusters, and porphyrins. This research can lead to mass-producible and economical spin photovoltaic devices.

## INTRODUCTION

The spin-photovoltaic cell is a brand-new field of interest where light absorption and power generation are dependent on magnetic properties of electrodes and spin properties of the electron. Spin-photovoltaic cells are exciting because new physics can be utilized to generate solar electricity. More importantly, spin-photovoltaic cells can be economically fabricated by the utilization of many earth-abundant materials, such as iron(Fe), nickel(Ni) and cobalt(Co). Initially, spin-photovoltaic effect was theoretically predicted around 1991 [1, 2]. The first category of spin-based photovoltaic cells focused on utilizing conventional charge-based p-n junction photovoltaic cell platform. In the theoretical studies, the conventional p-n junction was made spin sensitive by magnetic doping [3, 4] or by applying magnetic field [5]. The second category of spin-photovoltaic cells focus on the quantum transport. Many theoretical studies have predicted the spin-photovoltaic effect in quantum transport via quantum wires and nanoscale channels. [1, 6-8]. However, experimental efforts and progress have been insignificant in the above mentioned two categories of the spin- photovoltaic effects. Interestingly, many experimental studies have demonstrated the spin-photovoltaic effect in rather uncommon systems. Bottegoni et al. [9] utilized circularly polarized light to produce two spatially well-defined electron populations with opposite in-plane spin projections in the nonmagnetic semiconductor/metal system. The spatial separation of opposite spins was achieved by modulating the phase and amplitude of the light wavefronts entering in a germanium semiconductor layer covered with a patterned platinum metal overlayer. However, utilization of polarized light for observing spin-photovoltaic effect create potential issues with regards to practical applications of spin-based solar cells in direct sun radiation which is naturally unpolarized. Sun et al.[10] has utilized unpolarized light radiation to produce a spin-photovoltaic effect with a system of NiFe/$C_{60}$/AlOx/Cobalt(Co) based spin valves. In this work, the light was absorbed by the $C_{60}$ molecular film sandwiched between two ferromagnetic electrodes. This system yielded a spin- photovoltaic effect and was sensitive towards the direction of the magnetic moment of the ferromagnetic electrodes that were weakly coupled via the $C_{60}$ and AlOx layer.

In this paper, we demonstrate the spin-photovoltaic effect on molecule based spintronics devices that were produced by utilizing ferromagnetic metals like NiFe, and Co. We observed the spin-photovoltaic effect on a magnetic tunnel junction (MTJ) with exposed side edges (Fig. 1a) that was transformed into a molecular device (Fig.1b). In our case, paramagnetic molecules were covalently bonded with two ferromagnetic leads of the MTJ along the exposed edges (Fig.1b). The magnified view of a single molecule connection with two ferromagnets is shown in Fig. 1c. Our molecular device has the following three distinctions with respect to prior work by Sun et al. [10]: (a) Our molecular device is formed by covalently bonding molecular array across the tunnel barrier of Co/NiFe/AlOx/NiFe tunnel junctions along the edges (Fig. 1b). Whereas Sun et al. [10] sandwiched the $C_{60}$ molecules between Co and NiFe. (b) We have used a paramagnetic molecule that produced extremely strong exchange coupling between the ferromagnetic electrodes leading to significant change in magnetic properties and transport properties [11, 12]. Whereas in Sun et al. work [10] the magnetic coupling was supposedly insignificant because it was not discussed separately. We refer to our MTJ based molecular spintronics device (MSD) as MTJMSD in this paper. Here we report the observation of a photovoltaic effect on our MTJMSD using regular unpolarized white light at room temperature. We also report magnetic force microscopy(MFM), Kelvin probe atomic force microscopy (KPAFM), and Raman studies of the MTJMSDs to provide a comprehensive understanding and a potential mechanism behind the observed spin-dependent photovoltaic effect.

## EXPERIMENTAL DETAILS:

MTJ for the observation of photovoltaic effect were deposited on oxidized silicon with 300 nm silicon dioxide layer. MTJ thin film configurations, Ta(2-5nm)/Co(5-7nm)/NiFe(3-5nm)/AlOx(2 nm) /NiFe(10-12 nm) were produced by the liftoff method (Fig. 1d-h). This method involves depositing the bottom electrode on an insulating substrate (Fig. 1d) followed by the photolithography to produce a cavity in the photoresist to produce a cross junction (Fig. 1e). In the photoresist's cavity, ~2 nm AlOx (Fig. 1f) and top ferromagnetic electrode (Fig. 1g) were deposited. Liftoff of the photoresist produced a tunnel junction with exposed side edges (Fig. 1h). OMC was covalently bonded across the tunnel junction to form MTJMSD. Detailed MTJMSD fabrication protocol has been published elsewhere [13-17]. Typically, bottom and top electrodes were ~5 μm each and hence cross junction area was ~25 $\mu m^2$. The 3D atomic force microscope (AFM) image before treating an MTJ with OMC is shown in Fig. 1k. This MTJ remained intact after the interaction with OMCs (Fig. 1i); the physical height difference between the top and bottom ferromagnet did not change after the treatment with OMCs(Fig. 1j). OMC's thiol functional groups formed covalent bonding with the NiFe layers to produce molecular exchange coupling that dominated over the weak coupling via the AlOx tunneling barrier. This OMC coupling impacted the magnetic properties of the MTJMSDs.[11] An OMC possessed cyanide-bridged octametallic molecular cluster, $[(pzTp)Fe^{III}(CN)_3]_4[Ni^{II}(L)]_{4}$-$[O_3SCF_3]_4$ [(pzTp) = tetra(pyrazol-1-yl)borate; L = 1-S(acetyl)tris(pyrazolyl)decane] chemical structure and was paramagnetic in nature[18]. The HOMO-LUMO gap for the cubic core was theoretically calculated to be 0.33 eV; also the Ni and Fe atoms of each core were ferromagnetically coupled [19]. Spin-dependent photovoltaic effect was observed on at least ten MTJMSDs that were produced in four different batches. Transport studies of the MTJMSD were performed with a Keithley 2430 1kW pulse source meter and Keithley 6430 sub-Femtoamp source meter. White light radiation was supplied from a halogen lamp (Microlite FL 3000), The spectra of Microlite FL 3000 was characterized by Ocean Optics Spectrometers (Model S4000). We also used NaioFlex AFM, Picoscan AFM to conduct MFM. For KPAFM study we utilized Cr/gold coated cantilever that were produced by Budget Sensor®. For MFM study we utilized Co coated Nanoscience Nanosensor brand magnetized PPP-MFMR AFM cantilevers. Raman study was performed with JASCO NRS-4100 using 785 nm laser. During Raman study laser power, accumulation period, and exposure time were optimized.

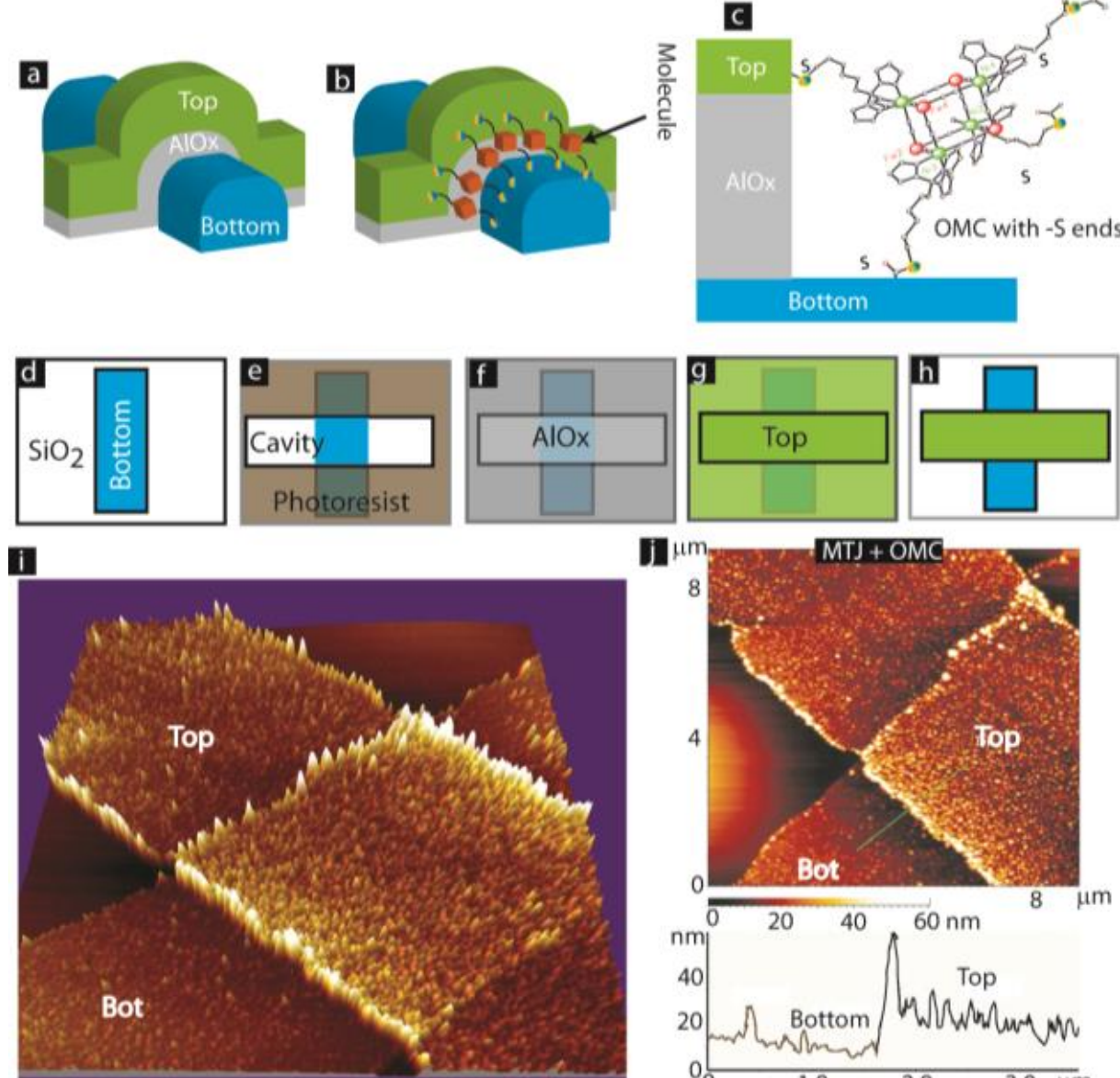

Fig.1: 3D view of an MTJ with exposed side edges (a) before and (b) after bridging molecular channels between ferromagnets. (c) Magnified view of OMC paramagnetic molecule covalently bonded to ferromagnets via thiol bonding. Fabrication of MTJMSD involve (d) bottom electrode deposition on insulating substrate, (e) photolithography for the deposition of (f) ~ 2 nm AlOx insulator and (g) top ferromagnet, followed by (h) liftoff. AFM of an MTJ (i) before and (j) after hosting hosting OMC channels.

## RESULTS AND DISCUSSION:

We investigated the photovoltaic effect on a wide range of tunnel junction based molecular devices with different metal electrodes and molecules [14, 17, 20]. However, the spin-dependent photovoltaic effect was only observed with MTJMSD utilizing Ta/Co(5-7nm)/NiFe(3-5 nm)/AlOx(2 nm)/NiFe(10 nm) MTJs and paramagnetic OMC. The Co/NiFe bilayer ferromagnetic films exhibited higher coercivity as compared to NiFe top electrode. Hence, even though OMC bonded with two NiFe layers present in the top and bottom electrodes but, the magnetic property of bottom NiFe layer was affected by the Co. The hysteresis loop for Co/NiFe was ~four time wider than that of top NiFe [11]. Ta only served as a seed layer to promote adhesion. A typical MTJ exhibited non-linear current-voltage graph indicating that ~2 nm AlOx insulating barrier produced expected tunneling type transport characteristics between two ferromagnetic electrodes (Fig. 2a). However, after the bridging of OMCs between two ferromagnetic electrodes of the MTJ, in the manner shown in Fig. 1c, effective MTJMSD transport varied from µA to pA range at room temperature (inset Fig. 2b). MTJMSDs' transport appeared in three broad categories that can be identified as high current state (in µA range at 50 mV, Fig. 2c), medium current state (nA range at 50 mV, Fig. 2d), and low current state (pA range at 50 mV, Fig. 2e). Out of these three current states medium and low current states, are termed as representative of suppressed current states; the magnitude of MTJMSD's current in medium (Fig. 2e) and low current states (Fig. 2e) were three to six orders of magnitude smaller than the magnitude of current observed at bare MTJ (Fig. 2a).

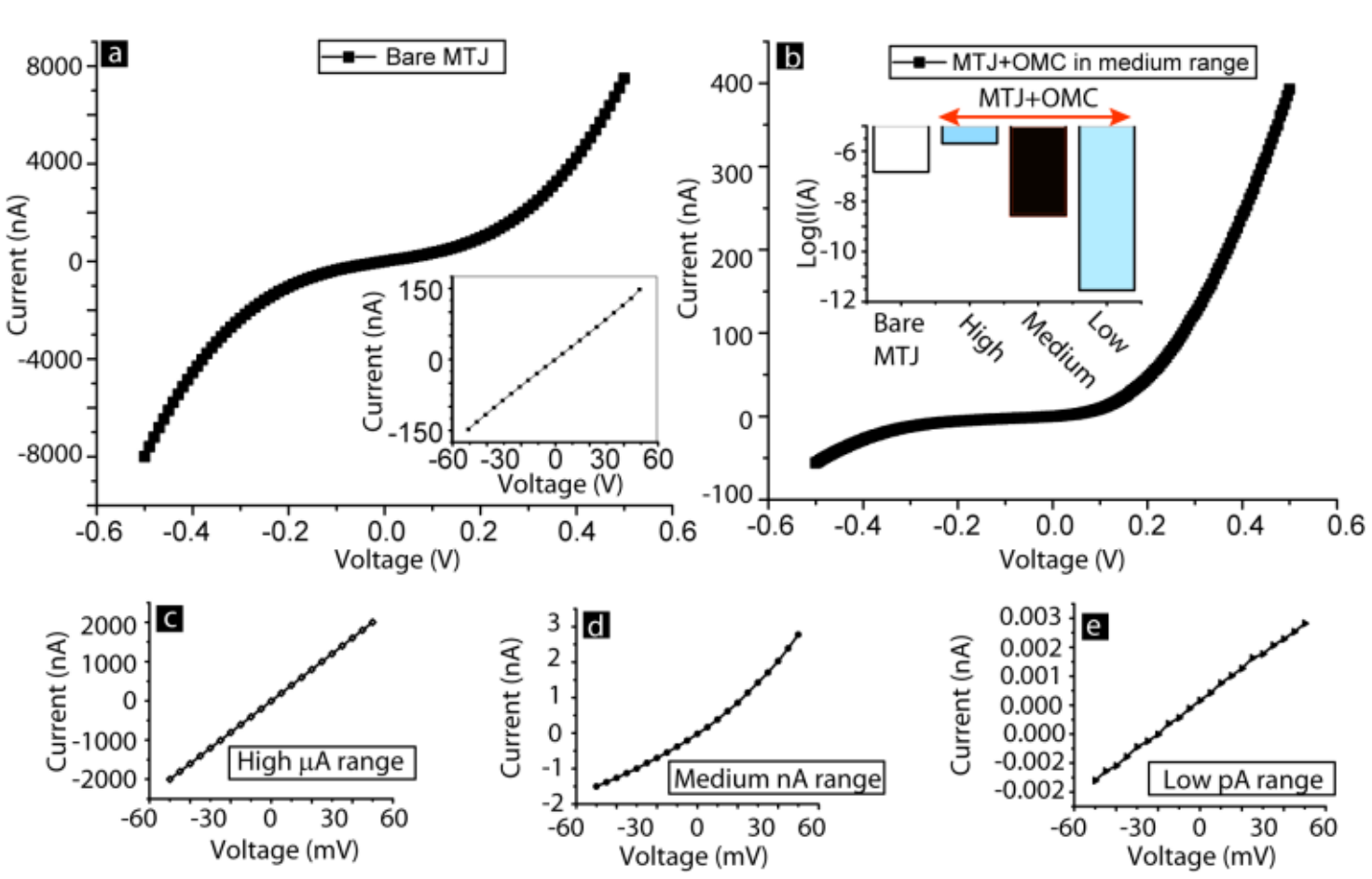

Fig. 2: Transport study of (a) bare MTJ showing tunneling type transport, inset show current magnitude for low voltage range. (b) After hosting OMCs as transport channels MTJMSD shifted among high low and medium current state, medium current state was the most stable state and showed diode like response. Low voltage current response of MTJMSD in (c) high, (d) medium, and (e) suppressed current state.

A bare MTJ by itself did not respond to light. It mainly responded to light after getting transformed in an MTJMSD and attaining a suppressed current state. In our prior publications, we have discussed the observation of current suppression [13, 20]. Here we briefly mention the current suppression in the context of spin-photovoltaic effect. We explained that such a large change in transport properties were associated with the OMC induced changes in the basic magnetic properties of the MTJs. We have also investigated and published a number of magnetic studies on MTJMSD[11, 12]. Magnetic studies were critical in showing that MTJMSDs' current suppression was associated with equally dramatic changes in the OMC induced magnetic properties [11]. Multiple magnetic studies gave direct evidence that OMCs created long-range changes in the magnetic properties of bare MTJs and the ferromagnetic leads used therein [11, 12]. OMCs produced very strong antiferromagnetic exchange coupling that impacted the ferromagnetic electrode at room temperature. OMC effect was confirmed by three independent magnetic measurement techniques, namely SQUID magnetometer, Ferromagnetic Resonance, and MFM. For the magnetic study, we utilized an array of ~7000 MTJ/samples to confirm the OMCs impact. Observation of OMC impact on ~7000 MTJ, with the same Ta/Co(5-7nm)/NiFe(3-5nm)/AlOx(2 nm)/NiFe(10 nm) thin film configuration as used in this paper, suggested that ferromagnetic films of the MTJMSD have attained new properties. However, in the previous work, we only discussed transport and magnetic properties[11-13, 20]. This paper is elaborating on the photovoltaic effect on the same MTJMSDs.

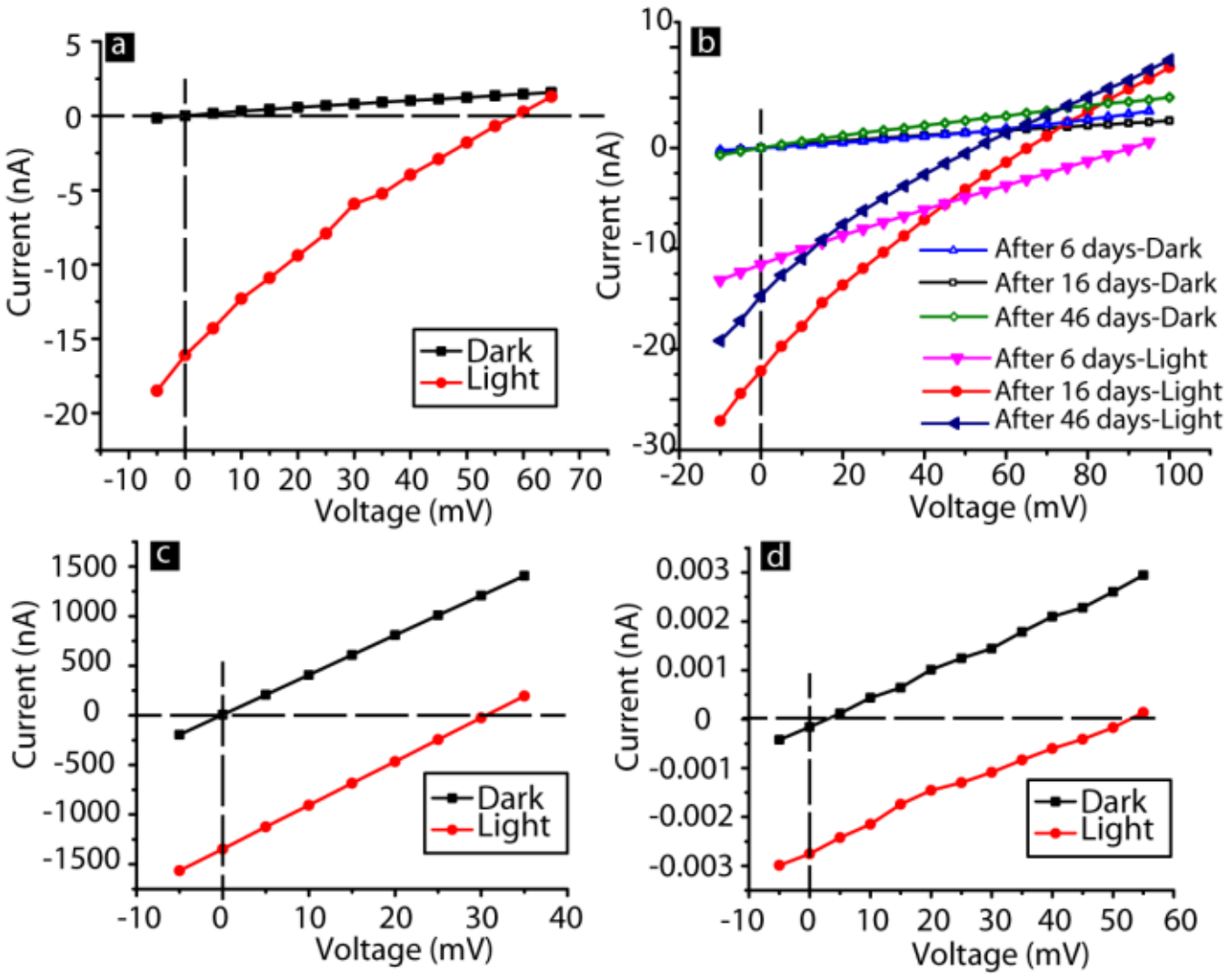

Fig. 3: MTJMSD's Photovoltaic effect in (a) middle current range. (b) Effect of time on photovoltaic response in middle current range MTJMSD's photovoltaic effect in (c) µA high current and (d) pA level suppressed current state. Suppress

We found that the medium current state, when MTJMSD was in the nA level current state, was more stable and produced the relatively stable photovoltaic response (Fig. 3a). Photovoltaic effect on the MTJMSD in the medium current range was time sensitive (Fig. 3b); however, this variation in photo-current was insignificant as compared to the magnitude of current variation between high (Fig. 3c) and suppressed current states (Fig. 3d). The photovoltaic response of the MTJMSD in the medium current state over 46 days showed that three measurements were still in 10-25 nA range (Fig. 3b). The open circuit voltage varied from 50-110 mV range. Such a time dependent change in MTJMSD's photovoltaic response is associated with the molecule induced magnetic ordering around the MTJMSD. We have discussed the impact of the MTJMSDs' magnetic orderings and their impact on transport properties elsewhere in this paper. We have also observed changes in MTJMSD transport and photoresponse in the high and low current states discussed in Fig. 3c-d. Both high and low current states ended up settling in the medium current level.

In the initial state, MTJMSDs' transport could be perturbed from medium nA current level to ~µA range by conducting multiple current-voltage studies or magnetoresistance study at a fixed voltage and by applying the magnetic field varying gradually up to ~300 Oe (Fig. 2). Such cases have been discussed in the prior work without a discussion on MTJMSDs' photoresponse, which is the focus of this paper [13, 20]. The photovoltaic effect of an MTJMSD that attained a temporary high current state is shown in Fig. 3c. On this MTJMSD sample, the high current state appeared after conducting multiple current-voltage studies. The representative unstable photoresponse in MTJMSD's high current state is shown in Fig. 3c. However, this high current state invariably changed to more stable several orders of magnitude lower suppressed current states; most frequently observed stable current state was a nA range medium current level. Typically, a transition from nA medium current level to pA level low current level occurred during the magnetization of MTJMSD under ~0.2 T permanent magnetic field [13, 20]. During magnetization, no current was forced through the device. In such cases, the photovoltaic effect produced ~pA photovoltaic current (Fig. 3d). Impact of magnetic field on MTJMSD's transport suggested that photovoltaic effect is dependent on the spin property of the electron. Due to the application of magnetic field photocurrent reduced by ~3 orders of magnitude (Fig. 3d) as compared to the stable nA level medium current state (Fig. 3c). More importantly, slight change in open circuit voltage (Voc) was observed. For the sample discussed here Voc changed from ~30 mV in the nA level medium current range to ~55 mV (Fig. 3d).

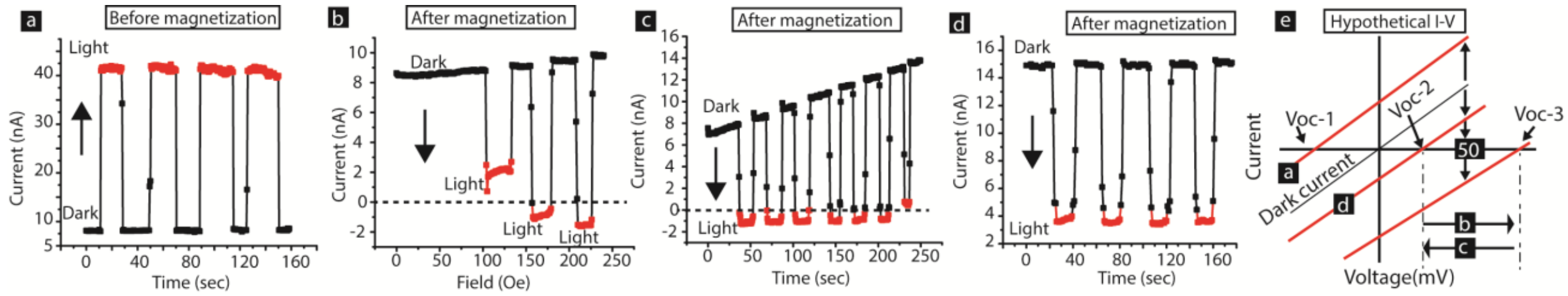

Fig. 4: MTJMSD's photovoltaic effect (a) before and (b-d) after magnetization. (b) MTJMSD's photovoltaic response under varying magnetic field. (c) MTJMSD's photovoltaic response without magnetic field 13 min after study shown in (b). (d) MTJMSD after 35 min of the study shown in panel (c). (e) Explanatory sketch for (a-d).

MTJMSD's photovoltaic effect exhibited higher sensitivity towards magnetic field in the initial state, or within first few days after MTJMSD formation. The response of light and magnetic field is discussed on an MTJMSD that was studied two days after the MTJMSD fabrication. Photoresponse was studied at 50 mV bias and by turning 96 $W/m^2$ white light on and off before and after MTJMSD magnetization by the ~0.2 T permanent magnetic field. Before the magnetization, this MTJMSD was in the nA level medium current range (Fig. 4a). This MTJMSD exhibited >4 folds current increase under light radiation (Fig. 4a). Interestingly, after the magnetization step the direction of photogenerated current reversed (Fig. 4b-d). Within 10 minutes after the magnetization by the permanent magnet, this sample was subjected to varying in plane magnetic field and photoresponse was recorded (Fig. 4b). Under the varying magnetic field, MTJMSD's photocurrent shifted from net positive to net negative. Such a change in the photocurrent sign suggests that the magnitude of open circuit voltage (Voc) was shifting. To generate net positive current Voc should be < 50 mV, at Voc=50 mV photocurrent is expected to be 0. When net photocurrent became negative Voc shifted to became more than 50 mV (Voc> 50 mV) (Fig. 4b). To investigate the stability of net negative photocurrent we studied photoresponse without varying magnetic field (Fig. 4c). Interestingly, with time dark current kept increasing at a much faster rate than photocurrent (Fig. 4c). After several minutes during the study photocurrent switched back to become net positive (Fig. 4c). We surmised that increase in dark current and change in photocurrent sign was due to equilibrium magnetic ordering of the OMC impacted ferromagnetic electrodes and should become rather stable after a period of relaxation. To check our hypothesis, we conducted photo response study after 35 minutes. As, expected MTJMSD exhibited stable dark current and a stable net positive photocurrent (Fig. 4d). It is also noteworthy that the starting current level of each study was close to the end point of prior study (Fig. 4a-d). We hypothesize that the nature of MTJMSD's current-voltage in light is changing (Fig. 4e). For Fig. 4a, increase in MTJMSD's current with light is only possible when photocurrent at V=0 intersect at positive current axis, and Voc is net negative (Fig. 4e). For identical experimental condition, under light radiation MTJMSD's current decreased below the dark current (Fig. 4b-d). These results signify that MTJMSD's current-voltage graph under light radiation has shifted to make saturation current (Isc) being negative and Voc net positive (Fig. 4e). During the study reported in Fig. 4b and Fig. 4c the magnitude of photocurrent is shifting between negative to positive sign. We surmise that this shifting sign of current is due to shifting Voc with regards to applied 50 mV bias. For data in Fig.4b, the Voc is moving from Voc<50 mV to Voc>50 mV(Fig. 4e). For data in Fig. 4c, the Voc is moving from Voc>50 mV to Voc<50 mV(Fig. 4e). After the initial upheaval an MTJMSD typically settled in a stable medium current state and exhibited a stable photo response for several days (Fig. 3b).

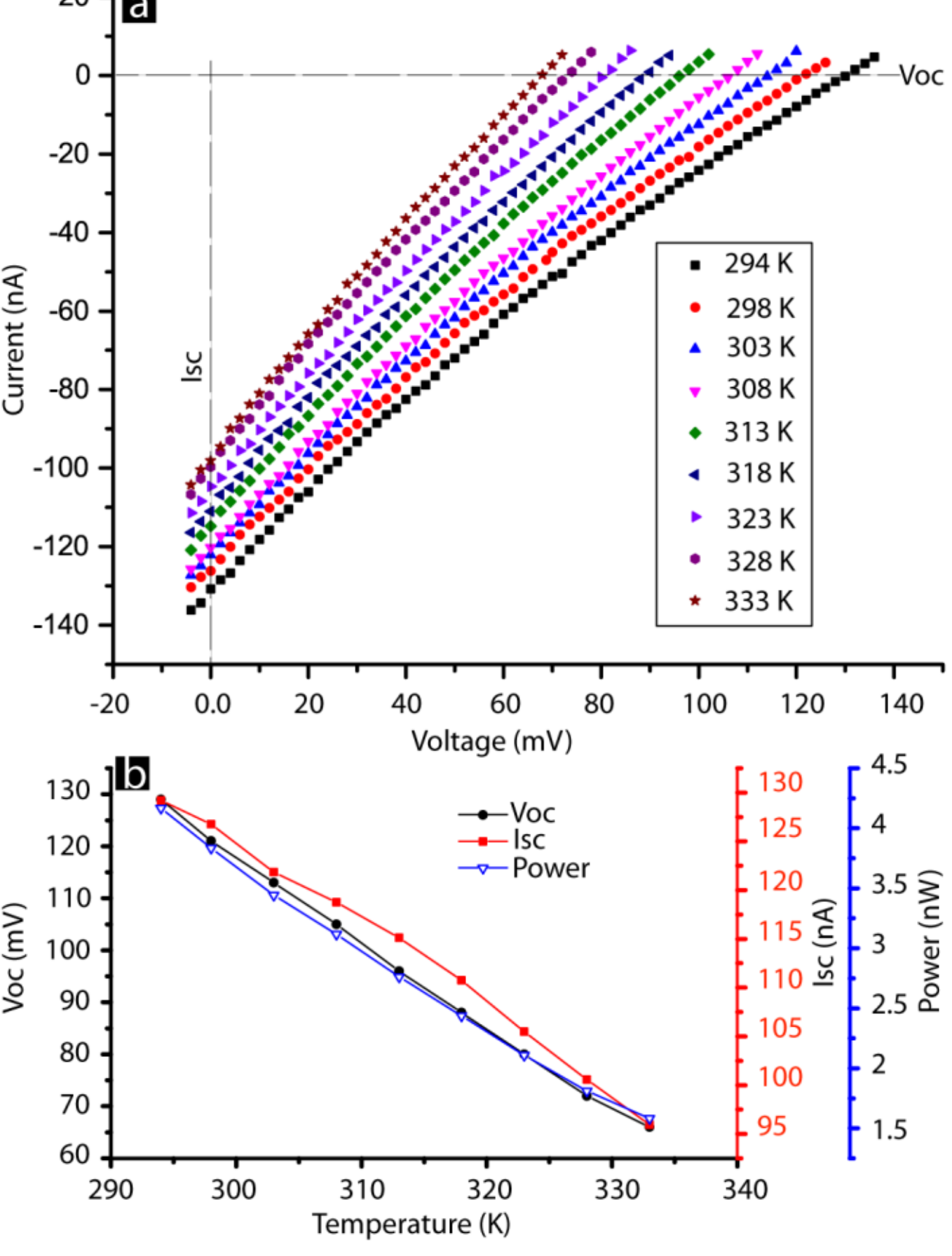

Fig. 5: (a) Effect of temperature on MTJMSD's transport. (b) Effect of temperature on open circuit voltage (Voc), saturation current (Isc), and power.

Besides, initial settling state we were unable to influence MTJMSD's photovoltaic effect with small magnetic field. We were unable to maneuver the direction of magnetization of the ferromagnetic leads in the MTJMSD's stable suppressed current state. The reason was that OMC produced an extremely strong exchange coupling between MTJMSD's magnetic lead that could not be overcome by the small magnetic field. As discussed elsewhere in this paper, three independent magnetic measurements confirmed that OMC produced very strong antiferromagnetic coupling between the ferromagnetic electrodes [11]. Application of 3T magnetic field was unable to overcome the OMC induced exchange coupling [11]. Also, experimental and theoretical studies showed that OMC induced exchange coupling strength was ~450 K range that is almost half of the inter-atomic exchange coupling strength [11]. Due to this strong OMC induced antiferromagnetic coupling ferromagnetic electrodes are expected to be aligned antiparallel to each other.

It is noteworthy that OMCs did not produce a photovoltaic effect on nanomagnetic tunnel junctions where nonmagnetic gold leads were utilized [21]. In addition, exposed tunnel junction edges where photo-active CuPc molecules were placed between two gold metal electrodes did not yield any photovoltaic effect or power generation [22]. Similarly, light-sensitive ~15 nm active molecular layer sandwiched between carbon electrodes did not produce measurable spin-dependent photovoltaic effect [23]. These studies suggests MTJMSD's photovoltaic effect is associated with magnetic materials and spin properties of the electrode.

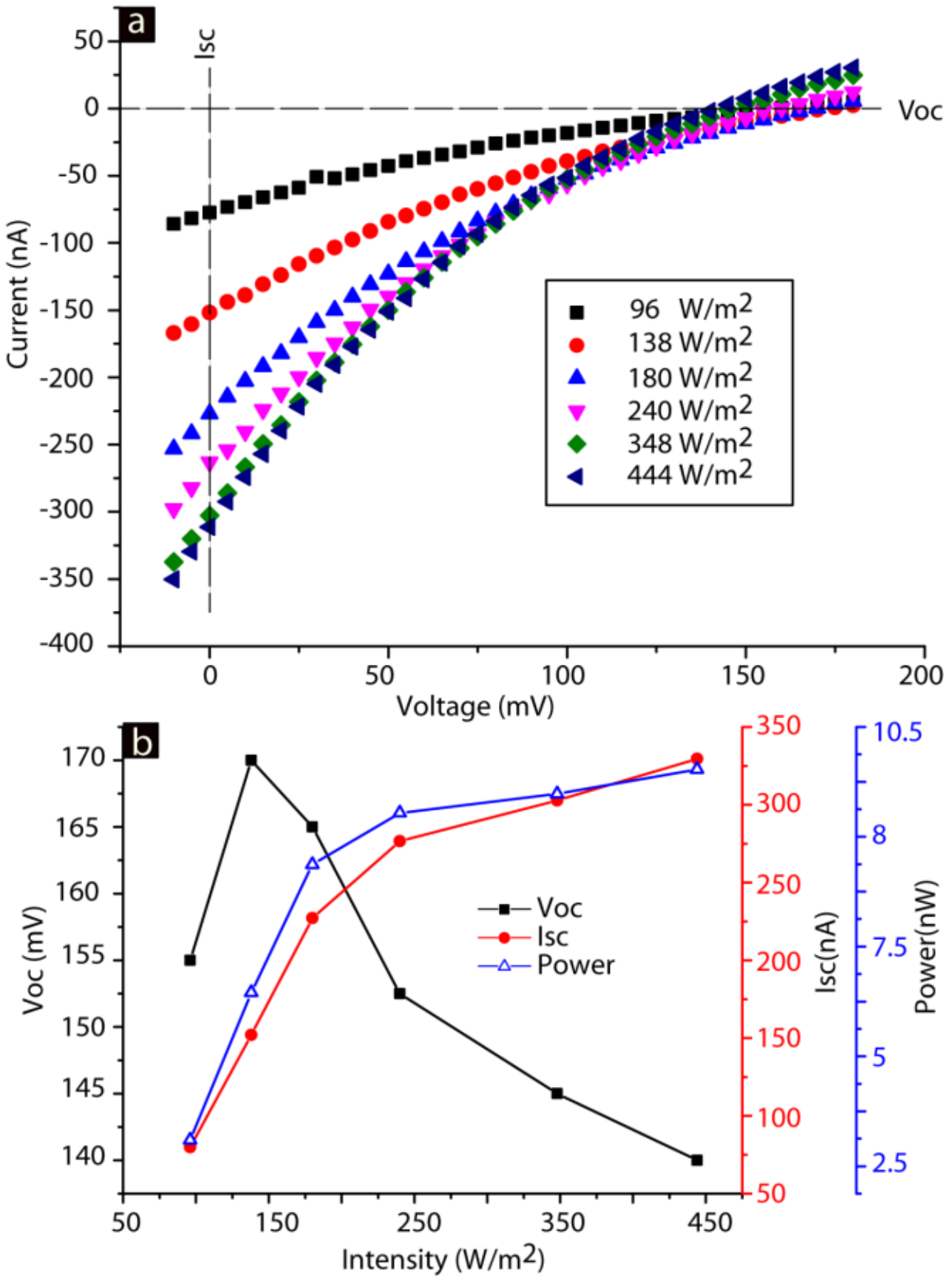

Fig. 6: (a) Effect of white light intensity on MTJMSD's transport. (b) Effect of light intensity on open circuit voltage (Voc), saturation circuit current (Isc), and power

We further investigated MTJMSD's photovoltaic effect under varying light intensity and temperature. Variation in temperature significantly impacted the photocurrent and open circuit voltage (Fig. 5a). For this study temperature varied from 294

K to 333 K range. This temperature range was selected to avoid the irreversible impact of temperature on magnetic leads and disturbance to MTJMSD's current state. It is noteworthy that magnetic electrodes start oxidizing at a high rate after ~360 K [15]. For this study light intensity was maintained at 96 $W/m^2$. Open circuit voltage (Voc) reduced from 129 mV to 66 mV linearly with temperature (Fig. 5b). On the other hand, the saturation current (Isc) under light radiation decreased linearly from 129 nA to 95 nA when temperature increased from 294 K to 333 K (Fig. 5b). For the calculation of power generated at MTJMSD we assumed ~0.25 fill factor. We calculated the power by multiplying fill factor, Isc, and Voc (Fig. 5b). Power generated at MTJMSD decreased linearly with temperature (Fig. 5b). We have discussed the efficiency aspect elsewhere in this paper. Power decreased by ~2.7 nW/K rate. This trend is like the one seen in the conventional p-n junction solar cells [24]. It is worth noting that MTJMSD's current in the darkness increased exponentially. The activation energy barrier was ~82±11 mV. Hence, the impact of temperature on MTJMSD was significantly different in the dark and light.

We also studied the effect of the light intensity on MTJMSD's photovoltaic effect (Fig. 6a). However, as compared to the effect of temperature the effect of light intensity was more prominent on saturation current at zero voltage (Isc) (Fig. 6a). We also estimated the amount of energy reaching on the MTJMSD cross section (Fig. 6a). The current-voltage graph for each light intensity is shown in the Fig. 6a. As light intensity increased the Voc first increased and then kept decreasing (Fig. 6b). Voc peaked around ~117 $W/cm^2$ (Fig. 6b). Whereas, Isc initially increased linearly with light intensity increasing from 96 to 139 $W/cm^2$ (Fig. 6b) and then tend to increase at a reduced rate. Like Fig.4c, we calculated power by assuming a fill factor (FF) of 0.25. The power is calculated to discuss the MTJMSD's energy conversion efficiency. For 444 $W/m^2$ light intensity the corresponding light intensity on an MTJMSD of 25 $\mu m^2$ area was 11 nW (Fig. 6b). However, the power generated at the MTJMSD was 11.5 nW. This estimation suggests that power generated from the junction is more than the incident light radiation power. This estimate suggested the photovoltaic power is generated from an area that is bigger than the typical tunnel junction area at the cross junction. To investigate this hypothesis, we investigated the OMC impact range on the MTJMSD. We mainly conducted magnetic force microscopy (MFM) on the MTJMSD cross junctions. MFM has been successful in capturing the OMC induced magnetic changes in pillars of several hundred MTJs of the similar thin film configurations [11].

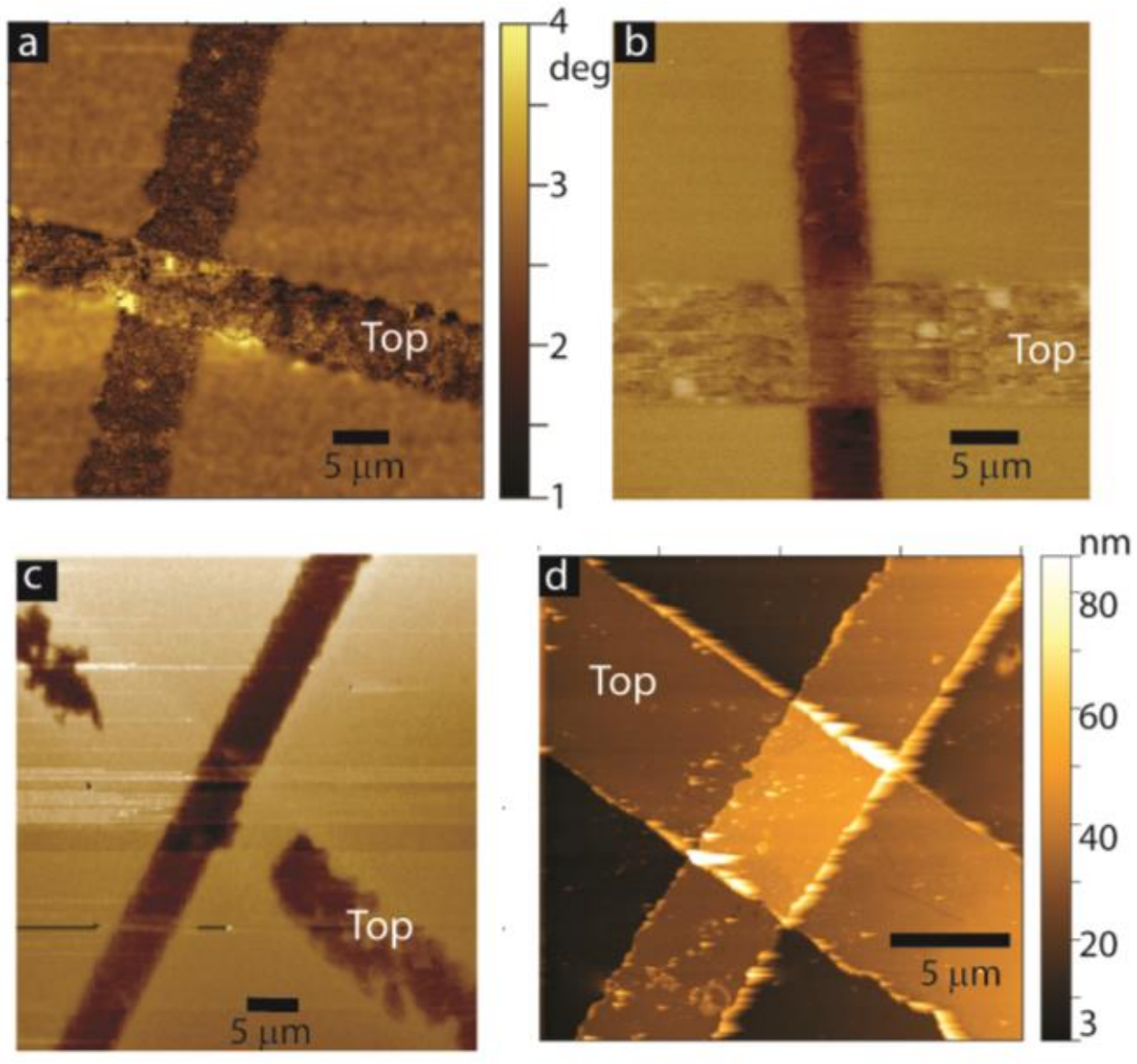

Fig. 7: Magnetic force microscopy (MFM) in (a) µA level high current state, (b) in low current state, and (c) in medium current state. Topographical AFM image of the junction shown in panel (c).

In the high current state, an MTJMSD exhibited same magnetic color (Fig. 7a). Also, MFM of the bare MTJ exhibited similar color for the top and bottom electrodes. However, in low pA level current state the top and bottom magnetic electrodes exhibited distinct color contrasts (Fig. 7b). Interestingly, in the nA level medium current state MFM study showed that a siginicant part of the top ferromagnetic layer lost magnetic contrast near junction vicinity (Fig. 7c). The loss of magnetic contrast around junction was not due to and physical damage of the top NiFe electrode. Topography image of the junction area in Fig.7c showed that top magnetic NiFe layer was physically intact (Fig. 7d). We observed that the disappearance of magnetic contrast near the MTJ junction area was a prevalent feature of the MTJMSDs reported in this paper. The OMC impacted regions near junctions are expected to be in an equilibrium state with the bulk of the ferromagnetic electrodes. We are unable to calculate the lower and upper bound of the OMC affected NiFe electrode regions. However, according to MFM image (Fig.7b-c) several tens of µm was affected by OMC beyond the junction area.

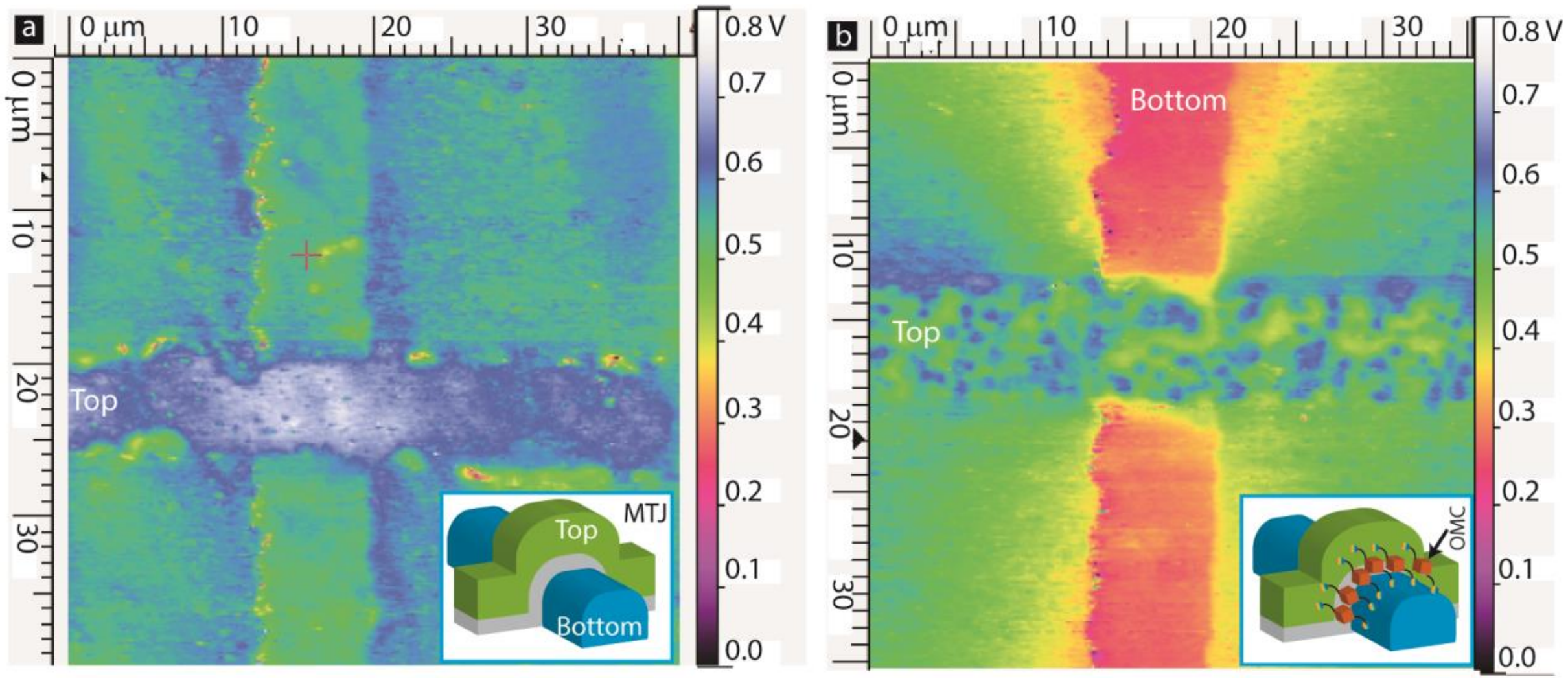

Fig. 8: KPAFM image of (a) MTJ and (b) MTJMSD.

To investigate the impact of OMCs on the MTJ junction area specific magnetic properties we studied an array of thousands of MTJs. An array of MTJ cylinders were transformed into MTJMSDs by bridging the OMC channels across the AlOx insulator. Topography imaging confirmed the physical presence of MTJMSD pillars. However, at the corresponding physical locations the MTJMSD's MFM contrast was negligible. We occasionally observed high MFM contrast at MTMSD sites. This high contrast MFM is believed to be an unaffected MTJ or failed MTJMSD, where OMCs did not produce strong coupling. An in-depth discussion about the MFM study on an array of MTJs is presented elsewhere [11]. In conclusion, the MFM studies provide direct evidence that OMCs were able to impact the large area of MTJs at room temperature. In the case of MTJMSD with cross patterns, OMCs are expected to impact ferromagnetic electrodes beyond the MTJ junction areas as observed in the MFM images shown in Fig. 7b-c. However, at this point we are unable to provide an exact estimate of OMC impacted ferromagnetic electrode areas responsible for the photovoltaic effect discussed in Fig. 3-6.

Our MFM studies indicated the OMCs produced a long-range effect on ferromagnetic electrodes (Fig. 7). We hypothesize that such large-scale OMC induced magnetic changes must also be observed in other types of experimental studies measuring different properties such as optical absorption and work function. To further confirm the OMC impact, we also conducted Kelvin Probe Atomic Force Microscopy (KPAFM) on MTJMSDs (Fig. 8). KPAFM of the bare MTJ, without OMCs showed a moderate difference of ~0.2 V in surface potentials of the top and bottom ferromagnets. However, an MTJ produced in the same batch showed a very different response after hosting OMC channels to become an MTJMSD. The bottom electrode's surface potential was ~0.4-0.6 V lower than the surface potential of the top ferromagnetic electrode. The large difference in surface potential is attributed to the OMC induced rearrangement of the density of electrons on the MTJMSD. KPAFM study also support the MFM studies that OMCs produce long range effect on the ferromagnetic electrodes' properties.

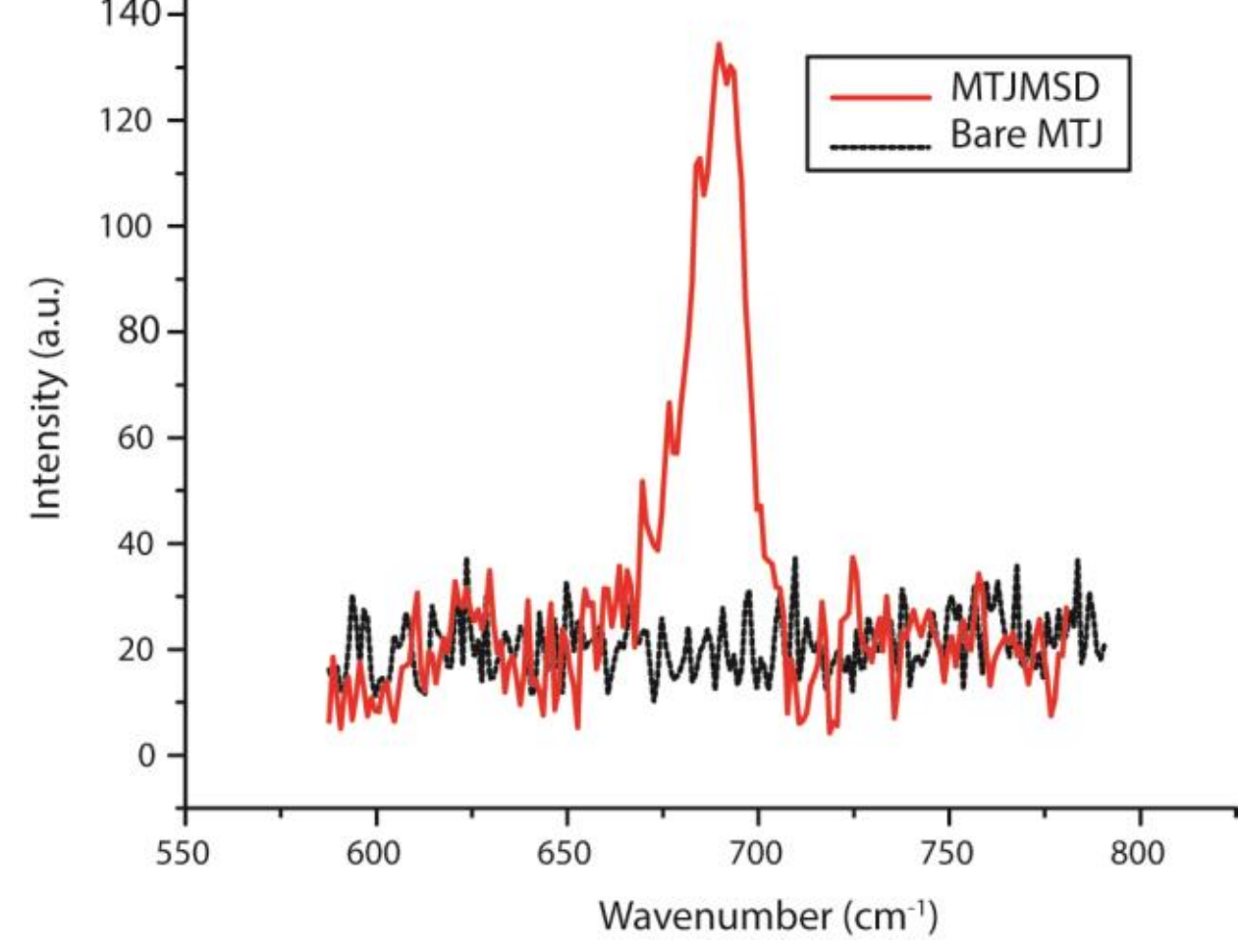

Fig. 9: RAMAN spectra of MTJMSD and bare MTJ

Photovoltaic effect is directly associated with the materials ability to respond to light energy. To investigate if our MTJMSD possessed a phase suitable for absorbing light radiation, we conducted Raman study using 785 nm laser on an MTJMSD and bare MTJ (Fig. 9). Raman spectra was recorded in the cross-junction area. We did not observe any noticeable signal in the Raman spectra for MTJ that we could attribute to the ferromagnetic leads or the tunnel junction (Fig. 9). However, MTJMSD's junction produced a prominent signal around 687 $cm^{-1}$ wavenumbers. This wavenumber corresponds to 14556 nm wavelength or ~85 meV energy. We do not believe this Raman response is due to any oxide formation because our MTJs were not heated beyond 95 °C. We demonstrated that on NiFe(80% Ni-20%Fe) acute oxidation start after 95 °C[15]. The Raman peak for the nickel oxide occurs around 590 $cm^{-1}$[25]. However, iron oxides are reported to exhibit multiple peaks; one of them was around 670 $cm^{-1}$ [26]. However, iron is only 20% of the NiFe and only surface may contain sub nm level iron oxide that is beyond the sensitivity range of Raman. Hence, it is not expected that peak we observed in the Raman spectra is associated with any oxide formation. Also, bare MTJ processed in parallel to the sample studied here did not show any peak under identical Raman experiment condition. Our Raman study provides direct evidence that the OMC impacted MTJMSD and made the junction responsive to the light radiation. Coincidently, the thermal activation energy barrier (~82±11 meV) calculated from the temperature dependent current -voltage studies in the medium current state was comparable to the order of light energy (~85 meV) corresponding to Raman peak. We surmise that MTJMSD possess an energy band gap of ~80 meV and hence capable of absorbing light radiation with energy ≥ the MTJMSD's band gap.

Two fundamental properties of a solar cell are that it should be able to absorb light to create the population of opposite charges/spins and subsequently separate them to generate current and voltage[24]. Raman study suggest that MTJMSD's can absorb light radiation energy and produce photogenerated spin or charges. Our KPAFM study suggest that MTJMSDs two metallic leads are at significantly different surface potential and may be associated with a viable mechanism to separate photogenerated spins to produce the observed photovoltaic effect in this paper. To explore the potential separation mechanism of opposite photogenerated spins we also calculated the barrier height($\phi$) and barrier thickness($s$) from the MTJMSD's transport data (Fig.10) and postulated a conceptual mechanistic model (Fig.11). The calculated barrier properties and conceptual model for the spin-photovoltaic effect can be adequately understood and appreciated by paying attention to MTJMSD's unique features that are remarkably different than those with the prior versions of molecular spintronics devices [16, 27].

We believe that spin-photovoltaic effect demonstrated here is due to the unique features of MTJMSDs. To make an MTJMSD nearly ~10,000 OMCs form molecular channels along the MTJ exposed edges [14]. It is noteworthy that only few nm scale electrical shorts can control the MTJ transport [28, 29]; chemically bonding ~10,000 OMCs along MTJ edges are suitable for creating strong molecular impact. As discussed elsewhere in this paper and shown in Fig 1c, a paramagnetic OMC's core makes bridge across the AlOx insulator with the help of two 10 carbon long alkane tethers, ending with thiol functional group. Thiol function group makes strong covalent bonding with the NiFe ferromagnet (Fig. 11a). It is noteworthy that on the NiFe surface, generally Ni persist in the elemental form [30]. Hence, when OMC interact with NiFe then NiFe's Ni atoms are expected to establish Ni-S bonds. The Ni-S bonds are thermodynamically more stable than Ni-O bonds [15], and hence OMC-NiFe bonds are expected to remain stable in the ambient conditions and create robust and reproducible interfaces on MTJMSDs. We could never break the Ni-S bond when attempted to reverse the OMC effect while keeping MTJMSD intact [14].

Also, OMC is endowed with useful attributes to produce observed molecule induced strong coupling and long-range effect. OMC's alkane molecule (tethers in Fig.11a) is far more superior spin channels as compared to oxide-based tunnel barriers, e.g. AlOx (Fig.11b vs. Fig.11c) [31]. Due to low spin-orbit coupling, and Zeeman splitting a 10 carbon long alkane molecule can preserve spin coherence for a long duration and length [31-34]. OMC's core connection to NiFe, via alkane tether and strong thiol-based metal-molecule coupling, produce unprecedented situation to transport ferromagnets' spin via the OMC's paramagnetic center with a net spin state. Due to geometrical constraints we are unable to measure the exact spin state of OMC when it is bridged between NiFe films (Fig. 11a). However, Park et. al. [19] has conducted Density functional theory (DFT) studies about the OMC's core, spin state, and energy levels. Importantly, OMC's $Fe^{3+}$ ions exhibit a low ground-state spin of S=1/2 due to strong ligand fields (Fig.11a). DFT study also indicated that each Fe ion in the core is ferromagnetically connected to the Ni ions via the -CN-ligand (Fig.11a) [19]. According to these DFT calculations HOMO-LUMO gap for the OMC core was 0.33 eV. Prior experimental study [18] indicated that OMC could attain as high as S=6 spin state; this study suggest that in the high spin state each Ni atom possess S=1 ground state (Fig.11a). OMC spin state is necessary to yield spin filtering effect. As long OMC maintain some spin state it can serve as a spin filter to only let a certain type of spin pass through it (Fig. 11c). As a result, OMC induced spin filtering modify the spin density of states and degree of spin polarizations on the MTJ's magnetic electrodes (Fig. 11d). Spin filtering has been discussed for many paramagnetic molecules [35, 36]. It is noteworthy that crystalline MgO tunnel barrier [37] has also shown spin filtering effect. A crystalline MgO based MTJ gave magnetoresistance performance, an attribute that depends on degree of spin polarization of ferromagnetic interfaces, almost ~10 time better than that of amorphous AlOx tunnel barrier based MTJs (Fig.11b) [38, 39]. It is apparent that ferromagnetic electrodes' degree of spin polarization is dependent on the spacer materials between the two ferromagnetic electrodes of an MTJ. In our MTJMSD we have two parallel spin channels, OMCs (Fig.11b) and AlOx (Fig. 11c). OMCs have advantageous internal molecular structure and well-defined Ni-S covalent bonding-based molecule-ferromagnet interfaces to supersede AlOx tunnel barrier in the MTJMSD (Fig.11a and d).

To investigate the potential mechanism, we analyzed the current-voltage data using Simmons transport model[17] and Brinkmann transport model[40] (Fig. 10). These two methods are commonly applied for the analysis of MTJ and molecular transport[41]. We estimated the effective barrier height($\phi$) and barrier thickness (s) on the bare MTJ and MTJMSDs and utilized to create hypothetical energy band diagrams to explain the spin-photovoltaic effect on MTJMSD (Fig. 11). For the MTJMSDs, we analyzed transport data in the high, medium, and low current state (Fig. 2). For the analysis junction area was equal to the AlOx tunnel barrier dimensions, 25 $\mu m^2$. We estimated that the molecular junction area is negligible as compared to the planar tunnel junction area. For the asymmetric medium current state (Fig. 2b), we separately fitted the Simmons model in the high current side (medium forward bias) and low current side (medium reverse bias). The summary of all the results is provided in Fig. 10.

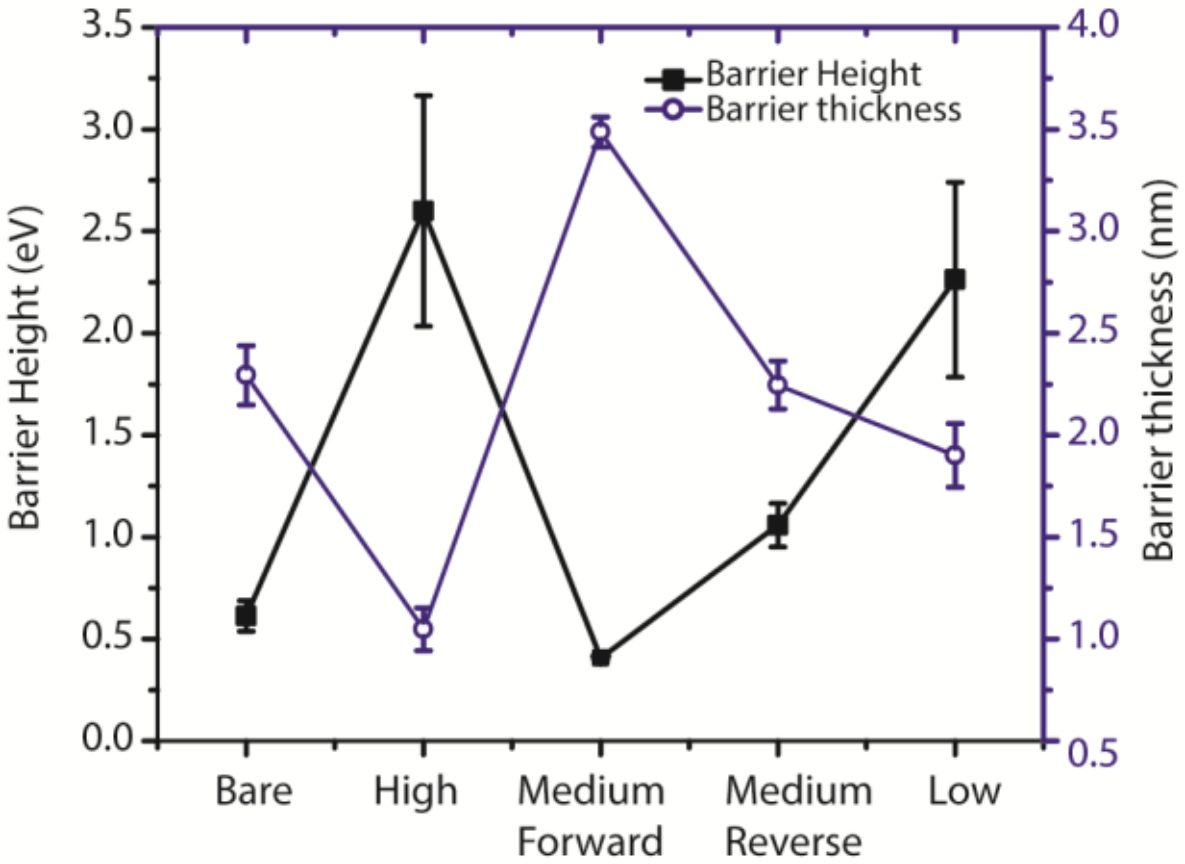

Fig. 10: MTJMSD's calculated barrier height and barrier thickness in various current states.

The barrier thickness of the bare MTJ was 2.3±0.15 nm (Fig. 10 and Fig.11b), and it was comparable with the physically measured AlOx thickness by the AFM [17]. Bare MTJ's barrier height was 0.6±0.08eV (Fig. 10 and Fig. 11b). However, transforming MTJ into MTJMSD resulted in a wide variation in the barrier height and barrier thickness. In the high current state, MTJMSD exhibited 1.1±0.1 nm barrier thickness. MTJMSD's barrier thickness in the high current state matches with the 10-carbon alkane tether molecule[41]. Alkane tethers with 10 carbon atoms were utilized in the OMC to bond with NiFe (Fig. 11a). This result hints that MTJMSD is dominated by the transport via OMC's core (Fig. 11e). However, barrier height in MTJMSD's high current state was 2.6±0.56 eV (Fig. 10). Interestingly, barrier height observed on a tunnel junction based molecular device with nonmagnetic leads was found to be ~1 eV [14] (Fig. 11c). We surmise that OMC induced modification (as graphically shown in Fig.11 c) in the ferromagnetic electrodes' surface potential or spin density of states led to 2.6±0.56 eV barrier height (as graphically shown in Fig.11 e). We surmise that in the high current state top and bottom ferromagnets are highly spinpolarized and aligned parallel to each other to let the highest possible current flow through the OMCs (Fig. 11e). A reflction of parallel alignment of the two FM electrodes is observable in the corresponding MFM image (Fig.7a).

In the MTJMSD's medium current state, barrier height and thickness was different for the forward (+ive voltage) and reverse bias (-ive voltage). An MTJMSD in medium-forward state exhibited 3.5±0.07 nm barrier thickness and 0.4±0.01 eV barrier height. This estimation hints that MTJMSD's transport in the medium forward state is influenced by a region that is neither AlOx (~ 2 nm) nor OMC's alkane tether

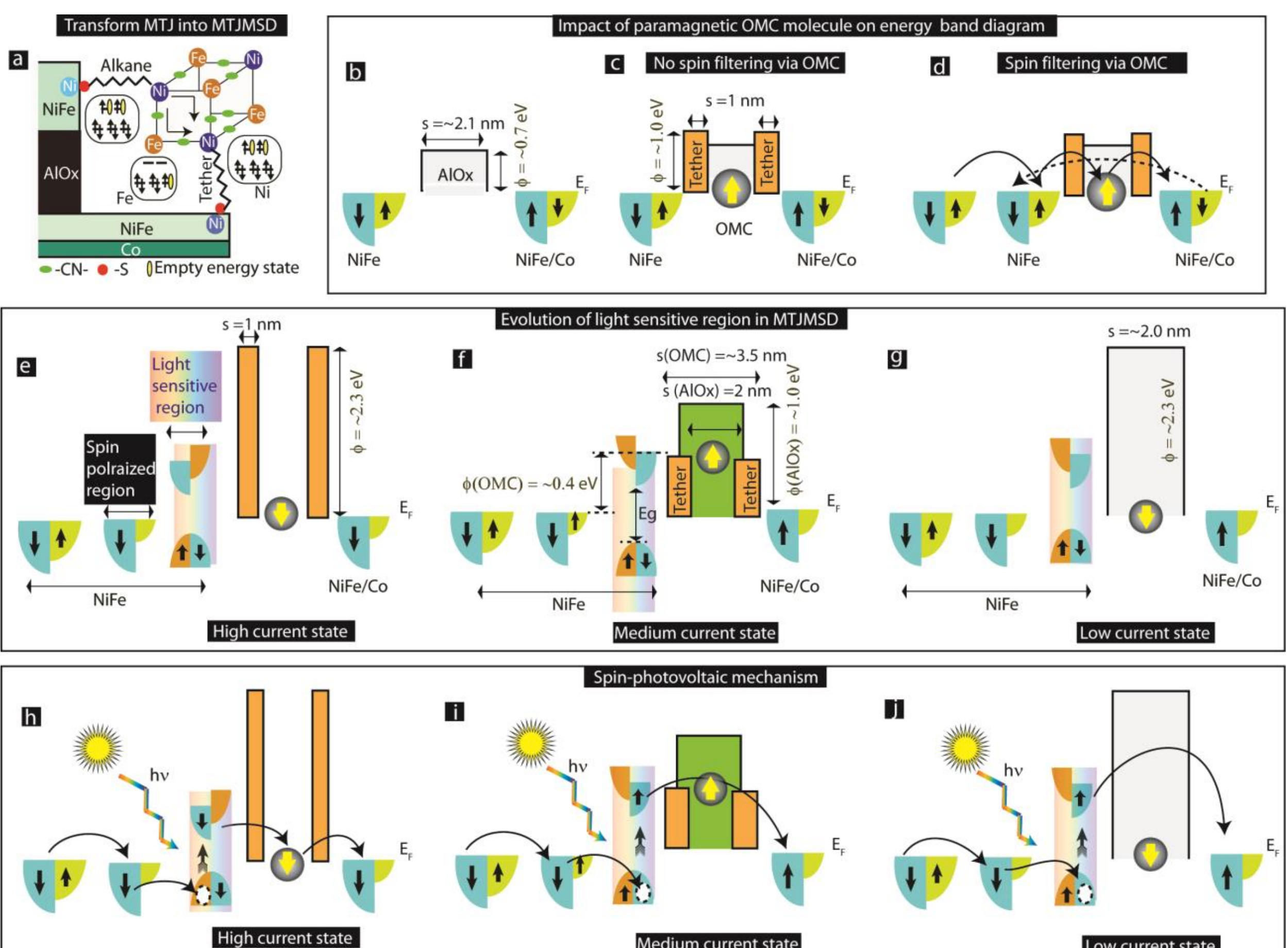

Fig. 11: (a) Conceptual diagram showing OMC bonding with ferromagnets, (b) Bare MTJ's energy band diagram, (c) Expected energy band diagram if OMC did not cause any spin filtering, (d) OMC leading to spin filtering and impacting ferromagnets. Evolution of light sensitive region on MTJMSD in (e) high current state when OMC is parallel to the two ferromagnets (fix barrier height), (f) medium current state, and (g) low current state. MTJMSD's photovoltaic effect in under illumination in (h) high, (i) medium, and (j) low current state.

(1 nm) (Fig. 10b). Barrier thickness of ~3.5 nm may correspond to the overall OMC's length (sum of two alkane tethers and octametallic cubic core, Fig.11a). Barrier thickness ~3.5 nm may also correspond to a combined region of AlOx tunnel barrier and OMC induced photosensitive ferromagnet layer. For both possibilities, the reduced barrier heights to ~0.4 eV level strongly suggest that metallic Fermi energy levels are not aligned with the OMC's energy level responsible for serving as intermediate state between the two ferromagnets (Fig. 11f). Coincidently, 0.4 eV barrier height is close to the 0.33 eV HOMO-LUMO gap calculated for OMCs from the DFT calculation [19]. We surmise that in the medium forward state OMC's energy level helping in the transport process is at 0.4 eV barrier height (Fig. 11f).

On the other hand, an MTJMSD in the medium-reverse bias state showed ~1.1±0.11 eV barrier height and ~2.2±0.12 nm barrier thickness. This barrier thickness of ~2.2 nm in the medium reverse bias state matches with the physical dimension of AlOx tunnel barrier used in our MTJMSD; for the bare MTJ AlOx, barrier thickness was ~2.3 nm (Fig. 10). However, barrier height in the medium reverse bias state (~1.1 eV) is almost double of the barrier height of the bare MTJ's AlOx tunnel barrier (~0.6 eV). The change in barrier height is expected due to the OMC's induced change on the ferromagnetic metals (Fig.7-9). This result suggests that in the medium current state transport is not occurring via OMC's core (in that case barrier height is expected to be ~1 nm, e.g. (Fig.11e), and OMCs have created asymmetric barrier heights by influencing the metallic leads (Fig. 11f). Hence, the difference in MTJMSD's barrier height for the reverse and forward bias condition suggest the presence of a built-in potential that may be operational in separating photo-generated charge carriers with opposite charges or spin, like the p-n junction case, and yield the observed photovoltaic effect.

In the low pA level, current state barrier height was 2.3±0.45 eV, and barrier thickness was 1.9±0.16 nm. Interestingly, barrier height in the pA level low current state (Fig.10 and Fig.11g) is comparable to barrier height in the µA level high current state (Fig.10 and Fig.11e). We surmise that OMC has transformed ferromagnets to become nearly 100% spin polarized and aligned antiferromagnetically (antiparallel magnetic moment) (Fig.11g). A reflection of almost opposite spin polarization of the two electrodes is seen in the MFM data (Fig. 7b). According to our Monte Carlo simulations, OMC's will align the two impacted OMCs antiparallel to each other, and in order to do so OMC has to be antiparallel to one of the magnetic electrode (Fig.11 g)[11]. In such case, transport is blocked via ~3 nm long molecule. However, transport is now dominant via the ~2 nm AlOx tunneling barrier that is present along with the OMCs's blocked channels (Fig.11g). Hence, our modeling matches with the conceptual picture (Fig.10g). In this state transport via molecule is prohibited and only ~ 2 nm AlOx tunnel barrier will be the viable conduction pathway.

Since MTJMSD exhibited a photovoltaic effect in all the three current states (Fig. 3) we hypothesize that OMC has created a photoresponsive magnetic layer in MTJMSD for each current state (Fig.11e-g). To realistically depict OMC interaction with the ferromagnetic electrodes in our conceptual diagram (Fig.11) we utilize the prior Monte Carlo simulations studies that we published for explaining experimental results[11]. According to our prior work[11], in the ground state OMC establish ferromagnetic coupling with one magnetic electrode and antiferromagnetic coupling with another magnetic electrode (Fig. 11f-g). Also, OMC induced exchange coupling between two ferromagnets was ~0.5 times of the Curie temperature of NiFe ferromagnet[11]. In the present case where the top and bottom electrodes extend beyond the junction the OMC induced exchange coupling may appear in multiple forms. Based on the MFM and KPAFM of the MTJMSD, the top ferromagnetic layer appears to develop a nonmagnetic region around the MTJ junction (Fig. 7c). This nonmagnetic area appears to be the result of striking a balance between two competing phases of the ferromagnetic electrodes. Around tunnel junction, OMC produces highly spin-polarized region(s) (Fig. 7b). The Fermi energy level of the OMC affected ferromagnets (Fig. 11e-g) is expected to be significantly different than that of bare MTJ (Fig.11b). Our KPAFM indicate that top and bottom ferromagnetic leads were significantly different from each other (Fig. 8). The change in surface potential indicates that relative difference between tunnel barrier, molecular energy level, and the resultant Fermi energy level has changed.

In the light, the photoactive region of the top electrode excites the spin up population that become spin down when staying in the excited state [2, 3] in all the states (Fig.11h-j). In the high current state of MTJMSD, excited spins move via the molecule to the bottom electrode (Fig.11h). Top electrode fills the spin vacancy in the shown manner (Fig. 11h). However, MTJMSD's high current state is unstable and transcend to a low current state. In the nA level medium current state photovoltaic effect was most prominent (Fig. 3-6). In this medium current state, we hypothesize that the light radiation excited the spin up electron that becomes spin down electron in the conduction band of the photoresponsive ferromagnetic region (Fig.11i). From the conduction band the spin down electron move over to the bottom electrode (Fig. 11i). OMC spin down state is expected to facilitate the transfer of photoexcited spin to the bottom electrode. The barrier height in the medium current state was in 0.4 to 1.0 eV range, indicating the ferromagnetic electrodes near junction are neither same as that with basre MTJ nor ~100% spin-polarized. In the case when magnetic electrodes are ~100% spin polarized, the barrier height is expected to be ~2.3 eV (Fig. 11g). Vacancy in the nonmagnetic area is expected to be filled by the spin-up electron of the NiFe electrodes (Fig.11 i). The spin photovoltaic effect in the MTJMSD's pA level current state is expected to be different. The pA level current state is due to the formation of one or two ferromagnetic electrodes becoming ~100% spin-polarized (Fig. 11g). In this state, the photoexcited electron cannot move via the molecule. Photoexcited spin up electron become spin-down and supposedly move via the AlOx tunnel barrier (Fig. 11j). The spin vacancy is filled by the spin up electrons from the top electrode. By this hypothetical mechanism an MTJMSD based spin photovoltaic exhibit power generation under light radiation.

By no means, we claim that our approach of utilizing tunneling models as a perfect method of investigating mechanistic insight behind the MTJMSD's spin-photovoltaic effect. However, it provides reasonable hints about the MTJMSD's state and OMC impact on barrier heights and thickness. Tunneling model does not incorporate the effect of temperature, magnetic anisotropies, and various forms of exchange couplings, e.g., biquadratic coupling and dipolar coupling. Hence, more accurate simulation and modeling are needed for better comprehension of spin-photovoltaic mechanism mentioned in this paper.

## CONCLUSIONS:

In this paper, we demonstrated the photovoltaic effect on MTJ based molecular spintronic devices (MTJMSDs). MTJMSDs exhibited three different current states termed as high (μA), medium (nA), and low (pA). In each state, light radiation produced the photovoltaic effect. OMC molecule appears to create robust exchange coupling between the two ferromagnetic electrodes of the magnetic tunnel junctions leading to significant changes in the electrical, magnetic, and optical properties of the ferromagnetic electrodes. OMC induced changes in the ferromagnetic electrodes also propagated outside the MTJ's perimeter. Magnetic studies, KPAFM, and Raman studies suggested that OMC transformed a ferromagnetic film into photo responsive material and produced a built-in potential in the MTJMSD. MTJMSD's ability to absorb white light radiation and ability to separate opposite spins in the three different current states lead to net photovoltaic effect. MTJMSD's photovoltaic response responded to the magnetic field. This paper mainly reports the experimental observations. Further investigation about the deeper understanding of the spin-photovoltaic effect is needed. We were also not able to provide an exact estimate of the energy conversion efficiency. It was experimentally challenging to determine the exact area responding to light radiation. Future work may focus on simultaneous KPAFM, MFM, and I-V measurements under dark and light for further understanding and new insights.

**ACKNOWLEDGEMENT:** Pawan Tyagi thanks Dr. Bruce Hinds and Department of Chemical and Materials engineering at University of Kentucky for facilitating experimental work on MTJMSD during his PhD. PT thanks Dr. Stephen Holmes and his postdoctoral scholar Dr. D.F. Li for producing OMC. An aged OMC batch was used for conducting Raman, UV-VIS, and KPAFM study. We thank Dr. Carlos Morillo of Jasco for the Raman study. KPAFM and MFM of MTJMSD in the high current state were completed by Christopher Riso. The preparation of this paper was in part supported by National Science Foundation-CREST Award (Contract # HRD- 1914751), and Air Force Office of Sponsored Research (Award #FA9550-13-1-0152). Any opinions, findings, and conclusions expressed in this paper are those of the author(s) and do not necessarily reflect the views of any funding agency and corresponding author's past and present affiliations.